\documentstyle[twocolumn,prl,aps,epsf,graphicx]{revtex}
\begin{document}
\draft
\twocolumn[
\hsize\textwidth\columnwidth\hsize\csname @twocolumnfalse\endcsname

\draft
\title{Off-Equilibrium Effective Temperature in Monatomic Lennard-Jones Glass}
\author{
        R.~Di Leonardo$^{1}$,
        L.~Angelani$^{2}$,
        G.~Parisi$^{3}$, and
        G.~Ruocco$^{1}$
        }
\address{
         $^1$
         Universit\'a di L'Aquila
         and INFM,
         I-67100, L'Aquila, Italy. \\
         $^2$
         Universit\'a di Trento and INFM, I-38050, Povo, Trento, Italy.\\
         $^3$
         Universit\'a di Roma {\rm La Sapienza}, INFN and INFM, I-00185,
	  Roma, Italy.
        }
\date{\today}
\maketitle
\begin{abstract}
The off-equilibrium dynamics of a monatomic Lennard-Jones glass is
investigated after sudden isothermal density jumps ({\it crunch})
from well equilibrated liquid configurations towards the glassy state.
The generalized fluctuation-dissipation relation has been studied and the
temperature dependence of the violation factor $m$ is found in agreement
with the one step replica symmetry breaking scenario, i.~e. at low temperature
$m(T)$ is found proportional to $T$ up to an off-equilibrium effective
temperature $T_{\!e\!f\!\!f}$, where $m(T_{\!e\!f\!\!f})$=1.
We report $T_{\!e\!f\!\!f}$  as a function of the density and compare it
with the glass transition temperatures $T_g$ as determined by
equilibrium calculations.

\end{abstract}
]


In the last few years the off-equilibrium dynamics of glassy systems
has been the object of intensive analytical and computational studies,
due to the very rich phenomenology exhibited and the possibility to obtain
detailed information on the
phase-space properties of the glassy system itself.
One of the most important features of the off-equilibrium regime is
the generalization of the fluctuation-dissipation relation \cite{BOUCH}.
At equilibrium the Fluctuation-Dissipation Theorem (FDT) \cite{PARSTA}
states that the susceptibility, $\chi_{AB}(t)$, that describes the response of
the variable $A(t)$ to an external field conjugate to the variable $B(t)$
is proportional, through the factor $\beta$=$( K_B T )^{-1}$,
to the derivative of the cross correlation function of the variables
$A(t)$ and $B(t)$ themselves, $C_{AB}(t)$. In the off-equilibrium case both
susceptibilities and correlation functions are two-times
quantities and the FDT no longer holds. However, as suggested
by Cugliandolo and Kurchan through analytical studies of soluble spin
glass models \cite{CUGLI}, the FDT can be generalized: the ratio between
$\chi_{AB}(t_1,t_2)$ and $\beta \dot C_{AB}(t_1,t_2)$ defines
a function, $X(t_1,t_2)$, that depends on the times $t_1,t_2$ only
through the actual value of the correlation function $C_{AB}(t_1,t_2)$,
$X(t_1,t_2)$$\equiv$$X(C_{AB}(t_1,t_2))$. The behavior of $X(C_{AB})$ 
gives information on the equilibrium order parameter in the low temperature
phase \cite{CUGLI,FRME,FRMEPAPE}, in agreement with numerical 
experiments on spin glasses \cite{MAPARIRUZU}. 

The conjecture of the similarity between structural glasses and
some spin glass model \cite{equiv} has been the basis to extend
these analytical results on spin glasses to structural glasses.
According to the latter conjecture one expects, in structural glasses,
that $X(C_{AB})$ is a two values function, with $X(C_{AB})$=1 at short
times and $X(C_{AB})$=$m \! < \! 1$ in the non trivial long time region
(the so called ${\it aging}$ region). The constant $m$ is expected to be
proportional to the temperature of the system up to an effective temperature
$T_{\!e\!f\!\!f}$, above which $m$=1 and, therefore, the FDT is recovered.
In the numerical tests of this scenario performed so far \cite{PABA},
the system is brought off-equilibrium at constant density by sudden
temperature jumps ({\it quench}), and the effective temperature
$T_{\!e\!f\!\!f}$ is found to coincide with the glass transition
temperature $T_g$. This is explained assuming that during
the {\it quench} the system "captures" the properties of the 3$N$-dimensional
potential energy surface at that temperature where the relaxation time
becomes comparable with the simulation time ($T_g$) and, at lower
$T$, it is no longer able to equilibrate. In this way the actual
aging dynamics coincides with those at $T_g$ even if the ``true'' temperature
is $T$$<$$T_g$. This is a reasonable scenario, but it is still unclear what
happens if the system is brought out of equilibrium without crossing the
thermodynamic state $(T_g,\rho)$.

In this Letter we numerically investigate the off-equilibrium
fluctuation-dissipation relation in a simple monoatomic glass,
whose particles interacts via a Lennard-Jones (LJ) potential
modified in such a way to avoid the crystalization occurring
in standard LJ. At variance with the usual constant density temperature
jumps method ({\it quench}), we use density jumps at constant temperature
({\it crunch}) in order to produce off-equilibrium configurations.
As expected, at low temperature, we find violation of the equilibrium FDT
in agreement with previous computational results in other model glasses
\cite{PABA}. We also find a linear temperature dependence of the
off-equilibrium parameter $m$ in the aging region, thus confirming the
conjecture of one step replica symmetry breaking scenario for real glasses
\cite{equiv}. More important, the effective temperature $T_{\!e\!f\!\!f}$,
below which $m$$<$1, is found very close to the glass transition temperature
$T_g$ at the density reached after the jump. While in the usual {\it quench}
from high $T$ at constant $\rho$ it seems natural to visualize the
process of freezing at $T_g$ during the quench, in our {\it crunch} method
the identification of the two temperature $T_{\!e\!f\!\!f}$ and $T_g$
evidences how the effective temperature depends only on the density of
the final state and does not depends neither on the initial state nor on
the path in $(T,\rho)$ plane. This observation gives a strong support to
the idea that the aging process can be considered as a pathway
in the 3$N$-d energy landscape from the off-equilibrium situation towards
the glassy minima\cite{bho}.

The investigated system is a Modified Lennard-Jones (MLJ) model glass,
with $N$=256 particles. The potential energy is $V$=$V_{LJ}$+$\delta V$,
where $V_{LJ}$ is the usual 6-12 Lennard-Jones interaction
(in the following all the dimensional quantities are expressed in
reduced units) and $\delta V$ is a many-body term that inhibits crystalization:
\begin{equation}
\delta V = \alpha \Sigma_{\vec q}  \; \;
\theta ( S({\vec q}) - S_0 ) \
\left [ S({\vec q}) - S_0 \right ]^2 \ .
\end{equation}
Here $S({\vec q})$ is the static structure factor and the
parameters in $\delta V$ ($\alpha^*$=0.8 and $S_0$=10) have been
tuned in order to avoid crystalization without
introducing a perturbation on the "true" LJ dynamics \cite{ADPR}.
The sum is made over all $\vec q$ with
$q_{max}$$-$$\Delta$$<$$|{\vec q}|$$<$$q_{max}$+$\Delta$, where
$q^*_{max}$=7.12$(\rho^*)^{1/3}$ and $\Delta^*$=0.34.
With this potential we can investigate all the phase space
without falling down in spatially ordered configurations
and with negligible corrections to the equation of state of monatomic LJ.

\begin{figure}[t]
\centering
\vspace{-1.cm}
\includegraphics[width=.35\textwidth,angle=-90]{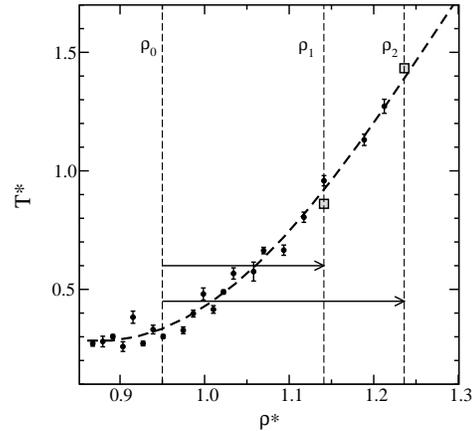}
\caption{The liquid-glass transition line in the plane ($T^*,\rho^*$) in reduced units.
The black points are determined from crossover in the temperature dependence
of equilibrium potential energy (the dashed line is a eyes guide).
The two grey squares are off-equilibrium effective temperatures $T_{\!e\!f\!\!f}$
determined from fluctuation-dissipation relation. Arrows qualitatively describe
the {\it crunch} process in the off-equilibrium analysis
(from $\rho^*_0$=0.95 to $\rho^*_1$=1.14  and to $\rho^*_2$=1.24).
}
\label{fig_2}
\end{figure}

First we study equilibrium properties of the system,
determining the liquid-glass transition line in the ($T,\rho$) plane.
The system is equilibrated at high temperature at different densities,
from $\rho^*$=0.87 to $\rho^*$=1.21, then it is slowly cooled
($dT^*/dt^*$=4.2$\cdot$10$^{-4}$) measuring the potential energy as a function of $T$.
\begin{figure}[t]
\centering
\includegraphics[width=.35\textwidth,angle=-90]{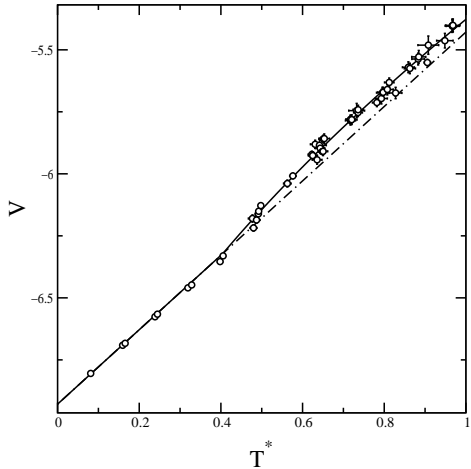}
\caption{
Potential energy vs temperature (all in LJ units) during the cooling
process from high temperature. The open circles are the molecular dynamics
data, the solid line is the fitted curve (see text) and the dashed line is
the harmonic extrapolation at high temperatures.}
\label{fig_1}
\end{figure}
In Fig.~\ref{fig_1} we report an example of the potential energy vs $T^*$
at density $\rho^*$=0.99.
It is evident the existence of a temperature $T^*_g$ that marks the transition
between two smooth regimes: at high $T$, in the liquid, as conjecture by Rosenfeld
and Tarazona \cite{ROTA} the potential energy follows a $T^{3/5}$ low while at low
$T$, in the glass, the linear behaviour expected for harmonic glasses is recovered.
In this way we are able to reconstruct the liquid-glass transition line
$T^*_g (\rho^*)$, as reported in Fig.~\ref{fig_2}. Below $\rho^*$=0.87 a
spinodal decomposition takes place, evidenced by the appearance of
``bubbles'' in the sample.

We focus now on the fluctuation-dissipation relation.
This relation connects the correlation function to the
response to an external field.
Let be $A(t)$ and $B(t)$ two generic  microscopic quantities,
and $\Delta {\cal H}$=$\lambda h(t) B(t)$
a perturbation added to the Hamiltonian, with $h(t)$=0 for $t$$<$$t_w$
and $\lambda$ an adimensional control parameter.
If we define the correlation function at
zero perturbation, $C_{AB}(t,t_w)$, as:
\begin{equation}
C_{AB}(t,t_w)=\langle A(t)B(t_w) \rangle_{_{\lambda =0}} \ ,
\label{corr}
\end{equation}
and the corresponding response function $\chi_{AB}$:
\begin{equation}
\chi_{AB}(t,t_w) = \lim_{\lambda \rightarrow 0} \frac{1}{\lambda}
\frac{\delta \langle A(t) \rangle_{_{\lambda}}}{\delta h(t_w)} \ ,
\label{resp}
\end{equation}
(hereafter the subscript $\lambda$ indicates that the average is
performed in presence of the perturbing term in the Hamiltonian)
then the equilibrium fluctuation-dissipation relation takes the form:
\begin{equation}
\chi_{AB}(t,t_w)=-\beta \  \partial C_{AB}(t,t_w) / \partial t_w \ .
\label{fluc_1}
\end{equation}
We notice that even if at equilibrium the presence of $t_w$ is redundant,
as $\chi_{AB}(t,t_w)$ and $C_{AB}(t,t_w)$ depend only on $t-t_w$,
we choose to use a two times formalism in order to easily generalize the above
formulas to off-equilibrium case.
Introducing the integrated response function $R_{AB}(t,t_w)$,
\begin{equation}
R_{AB} (t,t_w)= \makebox{$\int_{t_w}^{t}$} dt'\  \chi_{AB}(t,t') \ ,
\label{defQ}
\end{equation}
we obtain from eq.~(\ref{fluc_1}) the fluctuation-dissipation relation
in the integrated form:
\begin{equation}
R_{AB} (t,t_w)=\beta \left[ C_{AB}(t,t_w)-C_{AB}(t,t)\right]\ .
\label{fluc_2}
\end{equation}
In the present work we use the quantities $A(t)$ and $B(t)$:
\begin{eqnarray}
A(t) &=& \sqrt2\  (qN)^{-1} \Sigma_i \; \sin ( {\vec q}_i 
 \cdot {\vec r}_i(t) + \phi_i )  \ ,
\label{var_a}
\\
B(t) &=& - \sqrt2\   q^{-1} \Sigma_i \; \sin ( {\vec q}_i 
 \cdot {\vec r}_i(t) + \phi_i )  \ ,
\label{var_b}
\end{eqnarray}
where $\phi_i$ is a uniform random phase, ${\vec q}_i = q {\vec s}_i$,
${\vec s}_i$ a vector whose Cartesian components are random
variables $\pm 1$, and $q$=$2\pi n/L$, with $n$ an integer and $L$ the sample
size. Moreover we choose $h(t)=h_o\theta (t-t_w)$ ($h^*_o$=1 hereafter).
From Eq.~(\ref{corr}), after an average over the  random phases $\phi_i$,
one gets:
\begin{equation}
C_{AB}(t,t_w)= - (q^2N)^{-1}\Sigma_i \; \langle\ 
\cos ( {\vec q}_i \cdot \delta {\vec r}_i(t,t_w) )
\ \rangle_{_{\lambda=0}} \ ,
\label{corrAB}
\end{equation}
where $\delta {\vec r}_i (t,t_w)$=${\vec r}_i(t)$-${\vec r}_i(t_w)$, 
and, according to Eqs.~(\ref{resp}) and (\ref{defQ}),
\begin{eqnarray}
\nonumber
R_{AB}(t,t_w)\!&=&\!\sqrt2(q\lambda N)^{-1} \Sigma_i [ \ 
\langle\ \sin ( {\vec q}_i \cdot {\vec r}_i(t) + \phi_i ) 
\ \rangle_{_{\lambda}}  +
\\ 
&-&\ \langle\ \sin ( {\vec q}_i \cdot {\vec r}_i(t) + \phi_i )\ 
\rangle_{_{\lambda=0}}\! \ ] \ .
\label{respAB}
\end{eqnarray}
(Fortunately in the limit $N \to \infty$ the r.h.s of the 
equations becomes self-averaging).
A more suitable and easy to compute form is obtained 
in the limit ${\vec q}_i \cdot \delta {\vec r}_i \ll 1$.
In this limit, and averaging over ${\vec s}_i$,
Eq.~(\ref{corrAB}) becomes: 
\begin{equation}
C_{AB}(t,t_w)\!\simeq\! - q^{-2} + \Delta(t,t_w)/2 \ ,
\label{Climit} 
\end{equation}
having defined the mean square displacement:
\begin{equation}
\Delta (t,t_w) = N^{-1} \Sigma_i \;
\langle | {\vec r }_i(t) - {\vec r}_i(t_w) |^2 \rangle_{_{\lambda=0}}.
\end{equation}
Similarly, Eq.~(\ref{respAB}) becomes:
\begin{equation}
R_{AB}(t,t_w)\!\simeq\! (\lambda N)^{-1} \Sigma_i [ 
\langle {\vec f}_i \cdot {\vec r}_i(t) \rangle_{_{\lambda}}\!-\! 
\langle {\vec f}_i \cdot {\vec r}_i(t) \rangle_{_{\lambda=0}} ]
\label{Rlimit}
\end{equation}
where we have introduced  the force 
$\lambda{\vec f}_i$ acting on the particle $i$ due to the perturbation term 
in the Hamiltonian:
\begin{equation}
{\vec f}_i =\sqrt2  {\vec s}_i \ \cos ({\vec q}_i \cdot {\vec r}_i(t_w)+\phi_i) \ ,
\label{forze}
\end{equation}
this is a random variable due to the randomness of $\phi_i$ and ${\vec s}_i$.
Therefore, from Eqs. (\ref{Climit},\ref{Rlimit}), 
Eq. (\ref{fluc_2}) becomes   
\begin{equation}
2\ R(t,t_w) = \beta\  \Delta(t,t_w) \ .
\label{fdt_2}
\end{equation}

\begin{figure}[t]
\centering
\vspace{-.2cm}
\includegraphics[width=.35\textwidth]{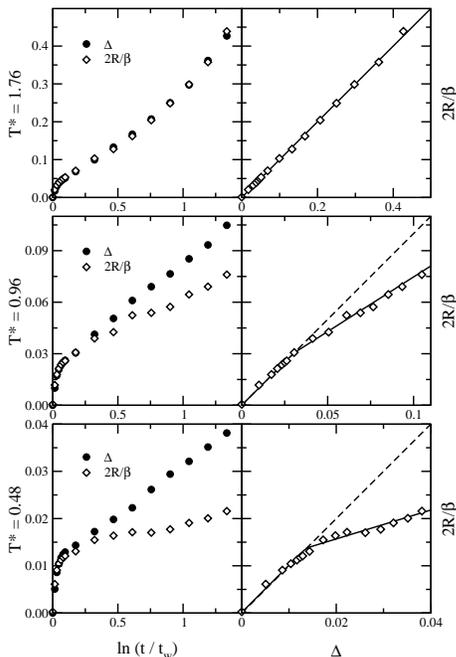}
\caption{Mean square displacement $\Delta$ and integrated response function
$2R/\beta$ in reduced units at $\rho^*_2$=$1.24$ for three temperatures
$T^*$=$0.48, 0.96, 1.76$.
The left side shows the $\log$ time dependence of the two quantities.
In the right side the response function vs $\Delta$. Dashed lines indicate
equilibrium fluctuation-dissipation relation, while full lines fit the
off-equilibrium aging region.
}
\label{fig_3}
\end{figure}

In the off-equilibrium case one can always define
a violation factor $X(t,t_w)$ in such a way to rewrite the
Eq.~(\ref{fluc_1}) as follows:
\begin{equation}
\chi_{AB}(t,t_w)=-\beta \ X_{AB}(t,t_w)\
{\partial C_{AB}(t,t_w)}/{\partial t_w} \ .
\end{equation}
In some recent papers \cite{CUGLI} it has been conjectured that,
in the aging region, $X_{AB}(t,t_w)$ depends on its arguments only through
the correlation function $C_{AB}(t,t_w)$. This allows to generalize the
fluctuation dissipation relation (\ref{fluc_2}) as:
\begin{equation}
R(C_{AB})= -\beta \  \makebox{$\int_{C_{AB}(t,t_w)}^{C_{AB}(t,t)}$} \
 dC_{AB} \ X(C_{AB}) \ ,
\end{equation}
or
\begin{equation}
{dR(C_{AB})}/{dC_{AB}}=\beta\ X(C_{AB}) \ .
\end{equation}
In our specific case we obtain:
\begin{equation}
2\ {dR(\Delta)}/{d\Delta}=\beta\ X(\Delta) \ , \label{gfdt}
\end{equation}
which generalizes Eq.~(\ref{fdt_2}).
In this work we have studied the generalized FDT, in the form expressed by
Eq.~(\ref{gfdt}). 
For sake of semplicity rather than perturbing the system
with the random force  ${\vec f}_i$ in Eq.~(\ref{forze}),
we have used a random force ${\vec f'}_i$ with the same variance of ${\vec f}_i$
($\langle f^2 \rangle=1$) and Cartesian components  $\pm1$ with
equal probability.
Equation (\ref{fdt_2}) has been numerically tested at equilibrium
(see the upper part of Fig. \ref{fig_3}).
  
The off-equilibrium states are obtained by {\it crunches} from an initial
equilibrated configuration at density $\rho^*_0$=0.95 at a given
temperature $T$ to a final state at the same $T$ and higher density
($\rho^*_1$=1.14 and $\rho^*_2$=1.24).
We set the time of the {\it crunch} as $t$=0.
For each investigated $T$ and $\rho$ we perform an isothermal molecular dynamics
simulations with $\lambda$=0 and calculate  $\Delta(t,t_w)$
as a function of $t$ and $t_w$.
Starting from the same $t$=0 configurations, after a waiting time $t_w$
we switch on the external field ($\lambda$$\neq$0) and measure the
response $R(t,t_w)$.
The strength of the external field $\lambda^*$=0.27 was chosen after
extensive tests of the linear dependence of the response on the perturbation.
All the measured quantities are averaged over $50$ initial configurations
and $10$ random extractions of the variables $\vec{f'}_i$.
The $t_w$ values investigated are $t_w^*$=1, 5, 10.
As an example,
in Fig. \ref{fig_3} we show the quantities $\Delta$ and $2R/\beta$ after a
{\it crunch} from $\rho^*_0$=0.95 to $\rho^*_2$=1.24 for three different
temperatures, $T^*$=1.76, 0.96, 0.48, and for $t^*_w$=5.
In the left hand part the two quantities are plotted versus $\log (t/t_w)$
while in the right side $2R/\beta$ is plotted as a function of $\Delta$.
All the data are compatible with the assumption that violation factor $X$
depends on $t$ and $t_w$ only through $\Delta(t,t_w)$, as verified comparing
the results obtained for different values of the waiting time.
Moreover the measured values $X(\Delta)$ are well represented by
a piecewise constant behavior ($X$=1 for short times and
$X$$<$1 at longer times) suggesting a one step replica symmetry
breaking scenario for structural glasses as conjectured in \cite{equiv}
and already found in \cite{PABA} in the cases of quenching systems of soft
spheres and LJ binary mixture.

\begin{figure}[t]
\centering
\vspace{-.8cm}
\includegraphics[width=.35\textwidth,angle=-90]{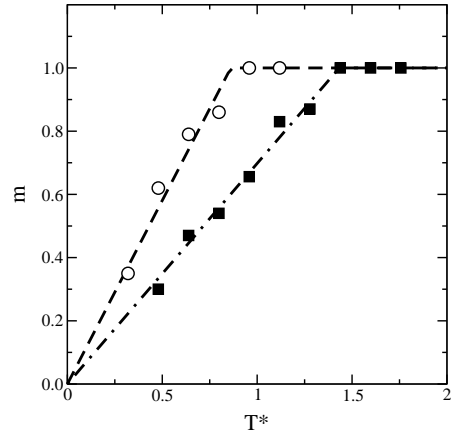}
\caption{
Temperature dependence of the violation factor in the off-equilibrium
aging region for two densities: $\rho^*_1$=$1.14$ (open circles) and
$\rho^*_2$=$1.24$ (full squares).
From the fits (dashed lines) one obtains:
$T_{\!e\!f\!\!f} (\rho^*_1)$=$0.86 $ and $T_{\!e\!f\!\!f} (\rho^*_2)$=$1.43$.
}

\label{fig_4}
\end{figure}

In Fig.~\ref{fig_4} we show the $T$ dependence of the violation factor
$m(T)$ in the aging region, for the two final densities analyzed.
It is evident a linear behavior up to a temperature $T_{\!e\!f\!\!f}$,
as found at low temperature in all know cases in which one step replica
symmetry holds. Fitting the data with the expression
$m(T)$=$a T \theta (T_{\!e\!f\!\!f}-T)$+$\theta (T-T_{\!e\!f\!\!f})$ we
extract the corresponding $T_{\!e\!f\!\!f}$ values, which turn out to
correspond with the previously measured equilibrium values of $T_g$
as evidenced Fig.~\ref{fig_1}. The coincidence between
$T_{\!e\!f\!\!f}$ at a given final density and $T_g$ at the same equilibrium
density evidences supports the hypothesis that, in the aging region,
the point representing the system in the 3$N$-d energy landscape
moves from the region pertaining to "high temperatures", where it has
been brought by the sudden density change, towards those configurations
where the characteristic time for structural rearrangement "diverges"
\cite{bho}.

In conclusion we have, for the first time, studied the off equilibrium dynamics
of a simple monatomic glassy system: a LJ potential where the crystalization
has been inhibited adding a small many-body term. The system has been brought
off-equilibrium via isothermal density jumps and the generalized
fluctuation-dissipation relation has been studied.
The found piecewise constant behavior of the the violation factor and its
linear temperature dependence agrees with one step replica symmetry scenario
conjectured for structural glasses.
For a given density, the emergent off-equilibrium effective temperature
$T_{\!e\!f\!\!f}$ seems to be very close the glass transition temperature $T_g$,
determined by equilibrium molecular dynamics measures.
Then, even if brought out of equilibrium by {\it crunches} at constant $T$
at final density $\rho$ without crossing the $T_g(\rho)$ transition point,
the system captures the properties of energy surface at $T_{\!e\!f\!\!f} \sim T_g (\rho)$.
The off-equilibrium effective temperature $T_{\!e\!f\!\!f}$ is then independent
on the particular path in $(T,\rho)$ plane crossing the transition line, and
depends only on the final reached density.


\end{document}